# Measuring the entropy of a neuron cell from its membrane current signal

Mahmut Akıllı[1]

1- Istanbul University, Institute of Science, Department of Physics, Istanbul, Turkey, akillimahmut@yahoo.com.tr

**Abstract**

The purpose of this study was to investigate how the entropy of a neuron cell can be measured using membrane ion current signals, which were recorded from neurons in the mouse medial prefrontal cortex (mPFC). The sample entropy and the Scalogram entropy were used as entropy measurement methods. It is well known that the entropy increases in the direction of the movement of the system towards the equilibrium. Therefore, in the process of the electrical activity of a living cell, the entropy is expected to reach a maximum at the moment when the membrane potential reaches the 'Nernst equilibrium potential' (or ionic equilibrium) of the ions. However, it was observed that the entropy values obtained by traditional calculations did not reach the peak at the equilibrium state of the ions. Therefore, two modifications to these measurement methods were proposed to adjust the entropy value to the maximum at the equilibrium potential of the ions. As a result of these proposed modifications, the entropy values were observed to peak around the equilibrium potential of the ions. These refined approaches were successfully validated using the Logistic map. Additionally, the entropy results were compared with Lyapunov exponents. The results show that the behaviour of living cells can be analysed using entropy measurements. The results also suggest that the method could be used to detect differences in the behaviour of tumour and normal cells, or the effects of drugs on cells.

## 1. Introduction

Entropy is a physical concept defined by Clausius and Boltzmann respectively and is known as the second law of thermodynamics. Systems in the universe maintain their existence by using energy. However, systems cannot convert all the energy they use into work. Here, they convert part of the energy they use into work and the remaining into heat. This remaining energy, which is converted into heat and not work, causes disorder in the system. The disorder of the system is also measured by entropy. In other words, the disorders caused by 'dead energy' that occur in energy-work conversion processes are called entropy. That is, entropy can also be defined as energy that loses its ability to do work and breaks the system. [1-5].

The entropy of closed physical systems left to spontaneous evolution increases with time and the entropy maximises towards thermodynamic equilibrium. However, living cells are non-equilibrium or open systems with flow, growth and change. Living cells also have lower entropy than their surroundings. Regarding this situation, Erwin Schrödinger, a famous physicist, published a book called 'What is Life' in 1944. In this book, he explained how the law of entropy works in living organisms. According to Erwin Schrödinger; if energy is not added to a closed system from the outside, entropy can constantly increase or remain constant. However, living organisms are open systems and receive energy from their surroundings in the form of nutrients or photons from the sun. Therefore, living systems have the ability to reduce their increasing entropy by exchanging either heat or matter with their environment. In this way, living organisms survive as long as they can keep their entropy low. [6]

Entropy is measured using methods based on probability distribution. The cell entropy can be estimated from the changing concentration distributions of the ions driven by the electrochemical gradients across the membrane. It is known that the entropy of a system increases as it moves towards equilibrium. Therefore, the entropy will be at its maximum at equilibrium. The ions move from a higher potential to a lower potential when the membrane of a living cell is stimulated at its 'resting potential'. Then, the cell entropy will be its maximum when the membrane potential is equal to the 'Nernst equilibrium potential' of the ions. Living cells have the ability to reduce their increased entropies by using free energy in the form of ATP.

In this study, the entropy of a single neuron cell was studied using ion membrane signals. For this study, membrane ion current signals from neurons in the medial prefrontal cortex (mPFC) of the mouse were used [7]. These ion current signals were recorded using Whole-Cell Patch Clamp technique. The accuracy of the results of the Sample entropy [8] and Scalogram entropy [9-11] used as measurement methods were checked by assuming that the cell entropy will be maximum at the Nernst equilibrium potential of the ions (or ionic equilibrium). However, it was determined that results consistent with this assumption were not obtained with the classical calculation method. To solve this problem, some modifications to the measurement methods were made to set the entropy value to peak at the Nerst equilibrium potential of the ionic currents. The entropy results were also supported by Lyapunov exponents. This methodology was also tested via Logistic map.

### 1.1. Electrical Activity of a Living Cell and its Entropy

A living cell often uses electrical signals to communicate and transmit information. Electrical activity is caused by the presence of ions such as sodium ($Na^{+1}$), potassium ($K^{+1}$), calcium ($Ca^{+2}$) and chlorine ($Cl^{-1}$) in the intracellular and extracellular media, as shown in **Figure** 1 (a). The cell membrane has an insulating feature against these ions, i.e. it does not allow the ions to pass through. As a result, ions accumulate on the interior and exterior surfaces of the cell membrane. An electrical potential difference occurs between the intracellular and extracellular as a result of the unequal distribution of ion concentrations. This is called the 'cell membrane potential'. Ions pass through the membrane via special channels called 'ion channels', which have closing and opening properties. Consequently, ion current consists of the movement of ions across the cell membrane driven by electrochemical gradients. [12-17]. Here, entropy can be defined as a measure of the dispersion of ions accelerated under the membrane potential [18]. This means that the amount of entropy is also a measure of the disorder distribution of the ions in the cell.

It is well known that systems always try to eliminate concentration gradients by maximising entropy. On the other hand, the systems have to use energy in order to reproduce the concentration gradients. For this purpose, living cells have proteins that act as pumps to move sodium and potassium or other ions from areas of low concentration to areas of high concentration. In this way, they recreate concentration gradients using free energy in the form of ATP. Thus, living cells try to survive by reducing their increasing entropy in this way. [6, 19-21]

After any electrical activity, cells return to their unique 'resting potential' state. Cells are negatively charged at the resting membrane potential. This means that the inside of a cell is more negative than the outside. Resting potential values vary according to cell type and species. [13-14]. The entropy of the cell can be considered as a minimum in the case of the resting potential. When ion channels open and positive ions flow into the intracellular, the potential difference between the intracellular and extracellular begins to decrease. In this case, the entropy of the cell also begins to increase. When the membrane potential and the Nernst equilibrium potential of the ions balance each other, i.e. when there is no net movement of ions across the membrane (or when the value of the ion current approaches zero, I≈0), the entropy is maximized. This is because, on average, ion transitions from the membrane occur at the same rate in both directions. The Nernst equilibrium potential expresses how much potential difference must exist between the two sides of the membrane to balance the migration of ions resulting from the concentration difference. This occurs when the electrical gradient balances the concentration gradient. On the other hand, as the negative ions begin to flow into the intracellular space and the positive ions begin to flow out of the extracellular space, a potential difference begins to form again between the intracellular and extracellular compartments. In this case, as the potential difference between the two sides of the membrane increases, the entropy of the cell also begins to decrease. Accordingly, as shown in **Figure** 1 (b), when an action potential occurs in a cell, cell entropy increases during the depolarisation phase and decreases during the repolarisation phase. [6, 19-21].

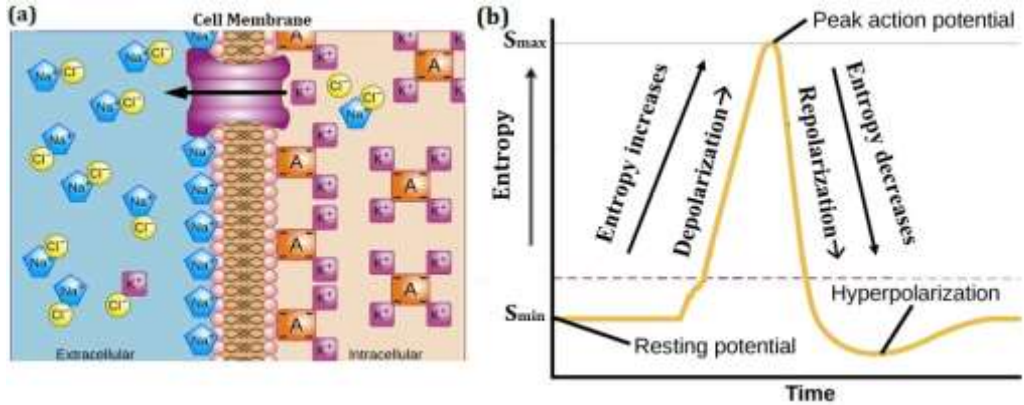

**Figure 1: (a)** The cell membrane current [22]. **(b)** Entropy change during action potential generation. The entropy of a living cell in a state of resting potential can be considered to be minimal ($S_{min}$). The entropy increases during the depolarisation phase. The entropy reaches a maximum ($S_{max}$) at the peak action potential. The entropy decreases during the repolarisation phase.

2. Methods
2.1. The Boltzmann-Gibbs-Shannon (BGS) Entropy

The concept of entropy ($S$) was first put forward by Rudolf Clausius in 1865 [23]. Macroscopically, he defined the entropy change ($\Delta S$) as the ratio of the change in heat energy of the system ($\Delta Q$) to its temperature ($T$), $\Delta S = \Delta Q/T$. Microscopic interpretation of entropy was made by Boltzmann. He demonstrated that entropy of a system in thermodynamic equilibrium is related to the logarithmic measure of the number of possible microstate ($W$), [24-26]

$$S = k_B \ln(W). \qquad (1)$$

Here, $k_B$ is the Boltzmann constant.

For a discrete random variable, the Boltzmann-Gibbs-Shannon (BGS) entropy is written in terms of probabilities as follows;

$$S_{BGS} = -k \sum_i p_i \ln p_i \,; \quad \sum_i p_i = 1 \qquad (2)$$

Where, $p_i$ is the probability of finding the system in a microstate $i$. If all the probabilities are equal, this definition (2) is equivalent to Boltzmann entropy (1). $k$ is a positive constant, used as $k = k_B$ in thermodynamics and $k = 1$ in information theory [27-28].

2.2. Scalogram Entropy

The scalogram entropy [9-11] is based on the scalogram S(s), which is the energy of the continuous wavelet transform (CWT) of a signal $f$ [29];

$$S(s) = \|\langle f, \psi_{u,s} \rangle\| = \|Wf(u,s)\| = \left(\int_{-\infty}^{+\infty} |Wf(u,s)|^2 du\right)^{1/2} \qquad (3)$$

Where, $Wf(u,s)$ is the CWT of a signal $f$, $\psi_{u,s}$ is a wavelet family scaling for time translating $u$ and scale dilating $s$ [..]. If the scalogram in equation (3) is rewritten for a finite time interval $I = [a,b]$;

$$S^{inner}(s) = \|Wf(u,s)\|_{J(s)} = \left(\int_{c(s)}^{d(s)}|Wf(u,s)|^2\, du\right)^{1/2} \quad (4)$$

Here $J(s) = [c(s), d(s)] \subseteq I$ is the maximal subrange in $I$. The inner scalogram $S^{inner}(s)$ provides information about the changes of energy densities at different scales of the signal over time. If the inner scalogram is normalized by time, average energy density distributions of the signal are obtained as independent of time. In this way, average energy densities at different scales can be compared within each other. The normalized inner scalogram $\bar{S}^{inner}(s)$ is obtained as follows [30-31]

$$\bar{S}^{inner}(s) = \frac{S^{inner}(s)}{(d(s)-c(s))^{\frac{1}{2}}} = (\overline{E_1}, \overline{E_2}, \overline{E_3}, \ldots, \overline{E_s}) \quad (5)$$

Where $\overline{E_s}$ represents the average energy of the signal for each scale $s$. The normalized inner scalogram (5) provides the average energy distributions of the signal at different levels of scale and does not include the time information. The probability distribution of the normalized inner scalogram that gives the probability of a certain scale is obtained as follows [9-11]

$$p_s = \frac{\bar{S}^{inner}(s)}{\sum_s \bar{S}^{inner}(s)} = \frac{\overline{E_s}}{\sum_s \overline{E_s}} = (p_1, p_2, p_3, \ldots, p_s); \quad \sum_s p_s = 1 \quad (6)$$

Where $p_s$ is the probability of scale $s$. The probability distribution (6) can also be written in the form of the Boltzmann distribution;

$$p'_s = \frac{e^{-\beta \overline{E_s}}}{\sum_s e^{-\beta \overline{E_s}}} \quad (7)$$

Where $\beta$ is a positive constant. The BGS entropy (2) can be calculated using the probability distribution in the equation (6) or (7). It can also be rewritten as;

$$S'_{BGS} = -k \sum_s p_s \ln p_s \quad (8)$$

Here, BGS entropy (8) is called 'scalogram entropy' because it is calculated from the normalized inner scalogram.

### 2.3. Sample Entropy

The sample entropy (SampEn) [8] was introduced by Richman and Moorman in 2000. Sample entropy is mathematical algorithm used to measure the degree of randomness of a series of data, like the approximate entropy (ApEn) [32-33]. Its algorithm is defined as the negative natural logarithm of the conditional probability between vectors $A^m(r)$ and $B^m(r)$;

$$SampEn(m, r, N) = -\ln \frac{A^m(r)}{B^m(r)} \quad (9)$$

Where, $m$ is length of the data segment being compared, $r$ is similarity (or tolerance) criterion, $N$ is length a series of data. $A^m(r)$ and $B^m(r)$, which are the total number of possible vectors, are obtained by calculating for each template vector as follows [8, 34];

$$B^m(r) = \frac{1}{N-m} \sum_{i=1}^{N-m} B_i^m(r) \quad (10a)$$

$$B_i^m(r) = \frac{1}{N-m-1} \sum_{j=1, j \neq i}^{N-m} [number\ of\ times\ that\ d[|x_m(j) - x_m(i)|] < r] \quad (10b)$$

$$A^m(r) = \frac{1}{N-m} \sum_{i=1}^{N-m} A_i^m(r) \quad (11a)$$

$$A_i^m(r) = \frac{1}{N-m-1} \sum_{j=1, j \neq i}^{N-m} [number\ of\ times\ that\ d[|x_{m+1}(j) - x_{m+1}(i)|] < r] \quad (11b)$$

In equations (10b) and (11b), the basic condition is that the distance function $d[x_m(j), x_m(i)]$ is smaller than the tolerance $r$. The tolerance is defined as follows, based on the standard deviation (std) of the data series;

$$r = c * std(data) \tag{12}$$

Where, $c$ is a coefficient usually chosen in the range of 0.1 to 0.25.

In this study, instead of the conventional tolerance (12), a new tolerance (or similarity) criterion (13) is proposed, defined as the ratio of the root mean square (rms) of the data series to its standard deviation (std);

$$r \equiv c * \frac{rms(data)}{std(data)} \tag{13}$$

Where, $c$ is a coefficient that can be chosen greater than zero, $c > 0$. If the tolerance $r$ is chosen according to equation (13), the sample entropy works well for membrane current analysis. Here, the tolerance coefficient $c$ is adjusted to maximise the cell entropy at the equilibrium potential of the ionic currents.

### 2.4. Lyapunov Exponent

The Lyapunov exponent $\lambda$ is a mathematical concept used to measure the degree of predictability of dynamical system behaviour. It is defined quantitatively as the average natural logarithmic divergence or convergence rate of nearby orbits in phase space. [35-36]. For mathematically defined systems $F(x_n)$, Lyapunov exponents are calculated using derivatives $F'(x_n)$ as follows

$$\lambda = \lim_{n \to \infty} \frac{1}{n} \sum_{i=0}^{n-1} \ln|F'(x_n)| \tag{14}$$

For calculating the largest Lyapunov exponent from a discrete random variable [37-38], let two nearby points in two time series $X_t$ and $Y_t$, the initial distance is $\Delta Z_0 = \|X_0 - Y_0\|$. Accordingly,

$$\lambda = \lim_{n \to \infty} \frac{1}{n} \sum_{i=1}^{n} \ln \left|\frac{\Delta Z_i}{\Delta Z_0}\right| \tag{15}$$

If $\lambda < 0$, dynamical system behaviour is predictable. However, If $\lambda > 0$, dynamical system behaviour is unpredictable, and the system is considered as chaotic. That is, the system is sensitive to a very small perturbation of initial state. [35-38].

### 2.5. Electrophysiological Recordings

In this study, membrane current signals that were recorded from L5 pyramidal neurons of the medial prefrontal cortex (mPFC) in mice were used [7]. The data were obtained from Dryad that is an open data publishing platform, freely accessible at https://doi.org/10.5061/dryad.66t1g1k2w [39]. Neuron membrane ionic current recordings of the mPFC were made by Whole-Cell Patch Clamp technique [40-42], using a voltage clamp protocol. **Figure 2** shows the current responses of three neurons evoked by +5 mV voltage steps (from –80 mV to –20 mV). Positive and negative signs show the direction of movement of the ionic current. The negative value of current indicates that the ionic currents are flowing into the cell. The positive value of current indicates that ionic currents are flowing out of the cell. The value of the current indicates zero at the Nernst equilibrium potential of the ionic current. **Figure 3** shows schematically the measurement of the entropy of a neuron cell from ion current signals.

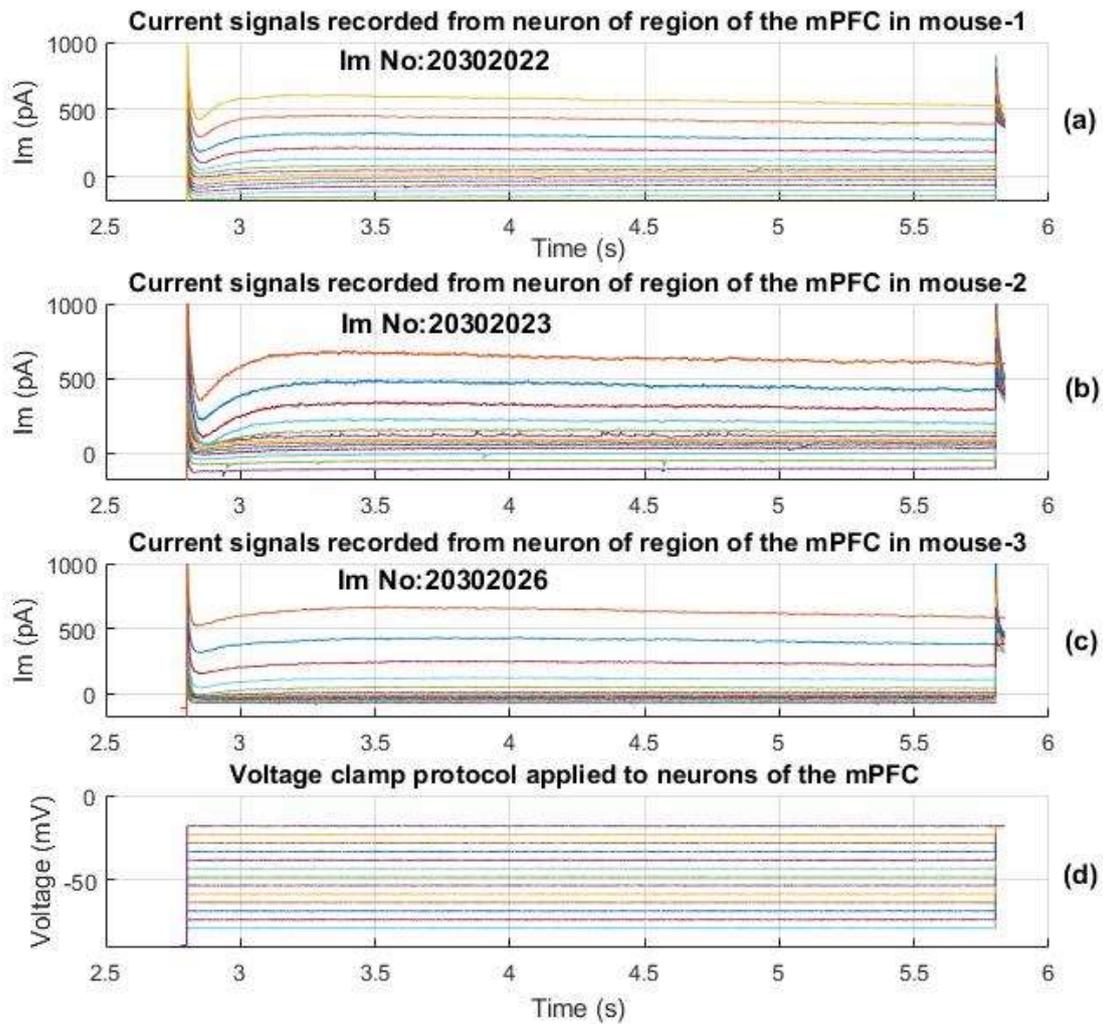

**Figure 2:** Ionic current recordings of L5 pyramidal neurons of the medial prefrontal cortex (mPFC). **(a)** Membrane current recordings of neuron of the mPFC number 20302022. **(b)** Membrane current recordings of neuron of the mPFC number 20302023. **(c)** Membrane current recordings of neuron of the mPFC number 20302026. **(d)** Voltage clamp protocol applied in the range of -80 mV to -20 mV in the Whole Cell Patch Clamp technique. [39]

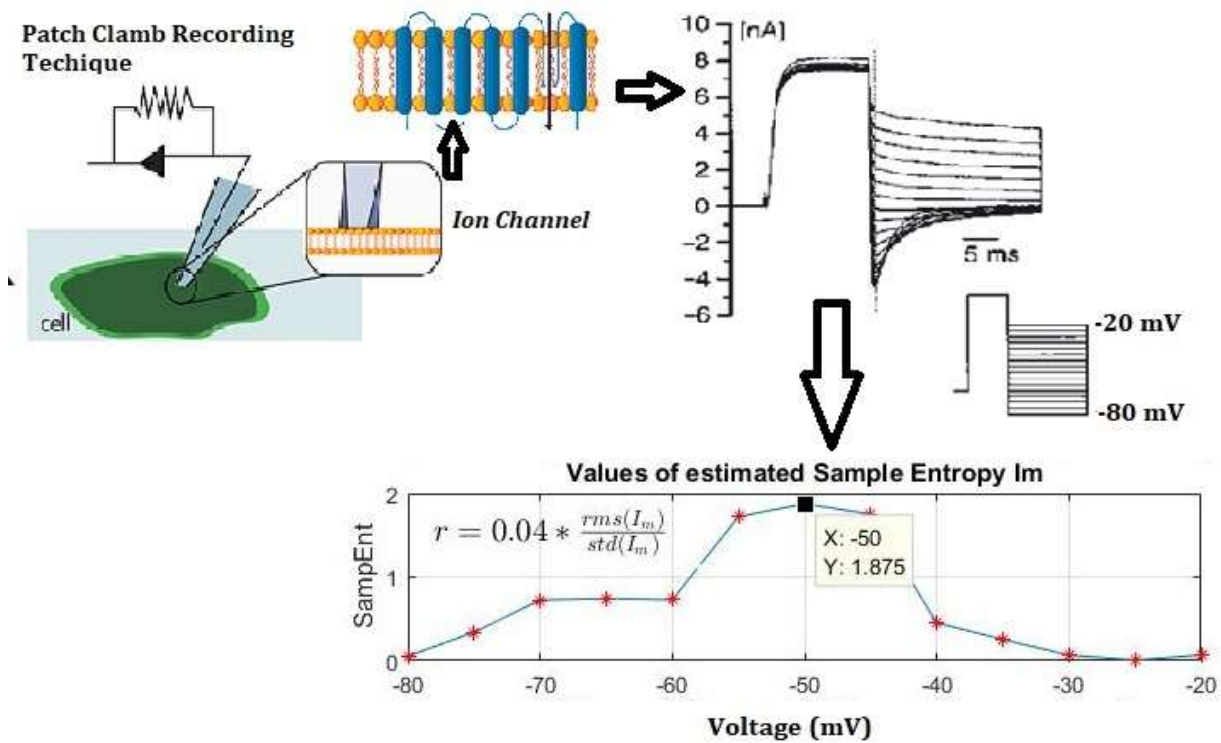

**Figure 3:** Graphical abstracts: Schematic diagram of the technical analysis of the entropy of a neuron cell measured from ion current signals.

### 3. Result
### 3.1. Entropy calculation for neurons in the mPFC

The entropy value is expected to reach a maximum at the ionic equilibrium stage on both sides of the membrane during the electrical activity process of a single living cell. That is, the entropy increases as the membrane voltage approaches to the Nernst equilibrium potential of the ions, and the entropy decreases as the membrane voltage moves away from their equilibrium. The value of the membrane current signal approaches zero around the Nernst equilibrium potential of ions. As demonstrated in **Figures** 2, 4(a), 7(a) and 8(a), the ionic current signals approach zero at potential values of -60 mV, -70 mV and -50 mV, respectively, for the three mPFC neurons. Entropy should peak at these voltage values. In this study, this information was used as a reference to check the accuracy of the calculations made for the membrane current signals. In accordance with this purpose; firstly, in the sample entropy calculations, the tolerance $r$ (13) is proposed as $r \equiv c * \frac{rms(data)}{std(data)}$. Secondly, the values of the scalogram entropy and the Lyapunov exponent were calculated for the inverse of the membrane current signal, where $I'_m = \frac{1}{I_m}$.

**Figure** 2 shows Whole-Cell Patch Clamp recordings of the three mPFC neurons used in this study. It shows the current responses recorded from neurons evoked in the range of –80 mV to –20 mV.

In the sample entropy calculations, $SampEn(m, r, N)$, the data length $N$ was chosen as 500 points. These points are taken from the 4th second of the current time series $I_m$ in **Figure** 2. The value of the embedding dimension $m$ was taken as 2. Generally the value of the tolerance $r$ is determined as $r = 0.2 * std(I_m)$.

**Figure** 4(c) shows the results of the sample entropy calculated for $r = 0.2 * std(I_m)$. Here, the sample entropy does not peak at -60 mV, which is the voltage value closest to the equilibrium potential of the ionic current. However, if the sample entropy is calculated for $r = 0.03 * \frac{rms(I_m)}{std(I_m)}$, the entropy peaks at -60 mV as seen in the **Figure** 4(b). For the other two neurons, as shown in **Figures** 7(c) and 8(c), the sample entropy peaks at the voltage values closest to the equilibrium potential of the ionic currents for $r = 0.03 * \frac{rms(I_m)}{std(I_m)}$ and $r = 0.04 * \frac{rms(I_m)}{std(I_m)}$, respectively.

In the scalogram entropy calculations, the data length was chosen as 1600 points. These points are taken from the 4th second of the current time series $I_m$ in **Figure** 2. The 'morlet' wavelet function with the scale range of $s_0 = 1$ and $s_1 = 256$ was used to compute the inner scalogram. The BGS entropy (8) was calculated from the probability distribution of the normalised inner scalogram (6).

**Figure** 5(c) shows the results of the scalogram entropy calculated for the current time series $I_m$. Here, the scalogram entropy does not peak at -60 mV, which is point closest to the equilibrium potential of ionic current. However, if the scalogram entropy is calculated for the inverse of the current time series $I'_m = \frac{1}{I_m}$, the entropy peaks at -60 mV as seen in the **Figure** 5(b). For inverse of the current time series of the other two neurons $I'_m = \frac{1}{I_m}$, as shown in **Figures** 7(d) and 8(d), the scalogram entropy peaks at the voltage values of -70 mV and -50 mV, which are closest to the equilibrium potential of the ionic currents.

Changes in Lyapunov exponents are expected to produce results parallel to changes in entropy. That is, when the entropy of a dynamical system increases, the value of the Lyapunov exponent is expected to increase accordingly.

In the largest Lyapunov exponent calculations, the data length was chosen as 500 points. These points are taken from the 4th second of the current time series $I_m$ in **Figure** 2. The algorithm improved (15) for estimating the largest Lyapunov exponents of experimental data was used.

**Figure** 6(c) shows the values of the largest Lyapunov exponent of the current time series $I_m$. Here, the largest Lyapunov exponent does not peak at -60 mV, which is point closest to the equilibrium potential of the ionic current. However, if the largest Lyapunov exponent is calculated for the inverse of the current time series $I'_m = \frac{1}{I_m}$, the Lyapunov exponent peaks at -60 mV as seen in the **Figure** 6(b). For the inverse of the current time series of the other two neurons $I'_m = \frac{1}{I_m}$, as shown in **Figures** 7(b) and 8(b), the largest Lyapunov exponent peaks at the voltage values of -70 mV and -50 mV, which are closest to the equilibrium potential of the ionic currents.

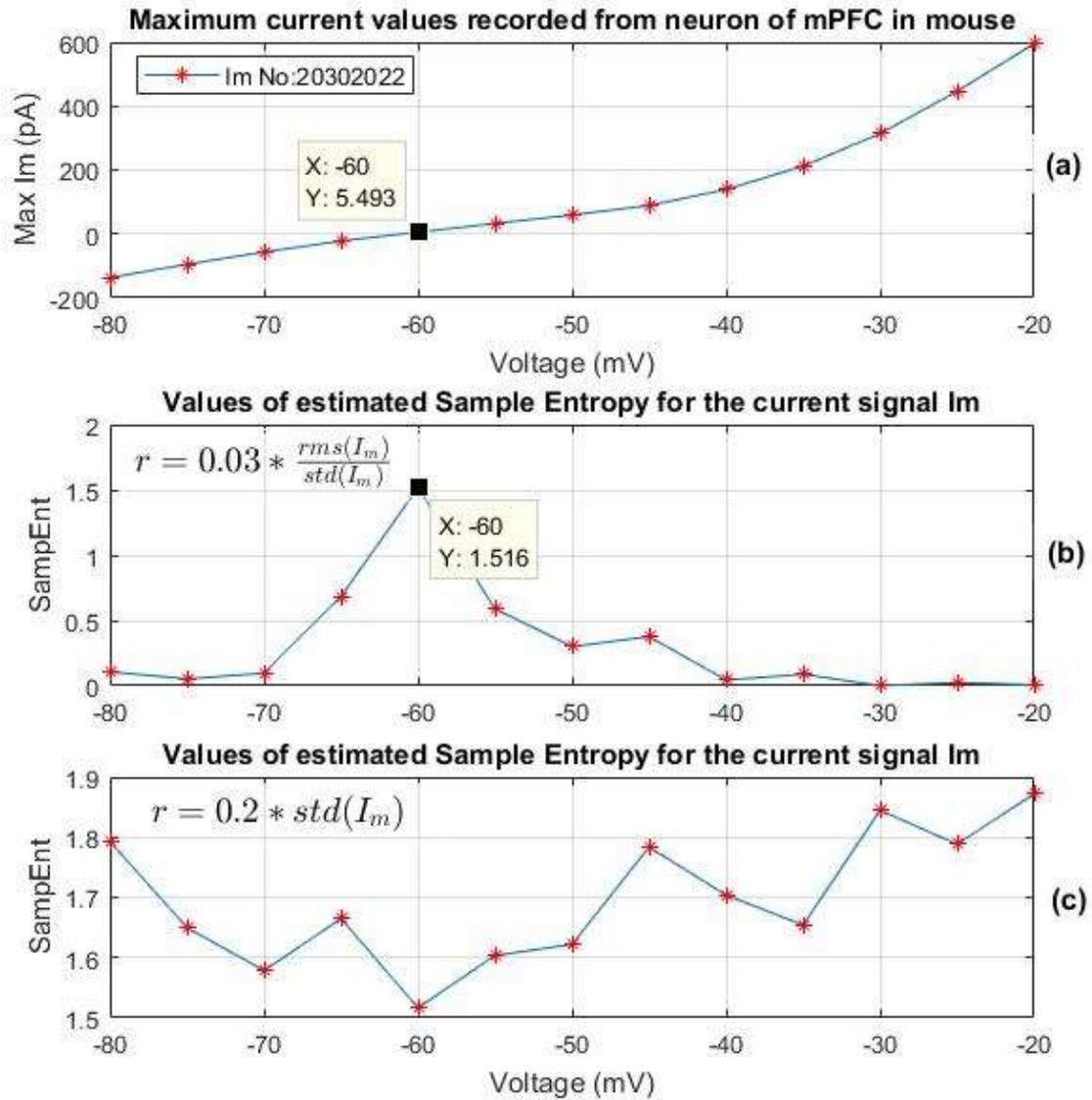

**Figure 4:** Registration file number: 20302022. **(a)** Maximum membrane current values recorded from neuron of the mPFC that were evoked in the range of –80 mV to –20 mV. The ionic current signal approaches a value of zero at a potential value of -60 mV (where the equilibrium potential of ions is noted). **(b)** The Sample entropy graph for membrane potential changes. Sample entropy values calculated for $r = 0.03 * \frac{rms(I_m)}{std(I_m)}$ **(c)** Sample entropy values calculated for $r = 0.2 * std(I_m)$.

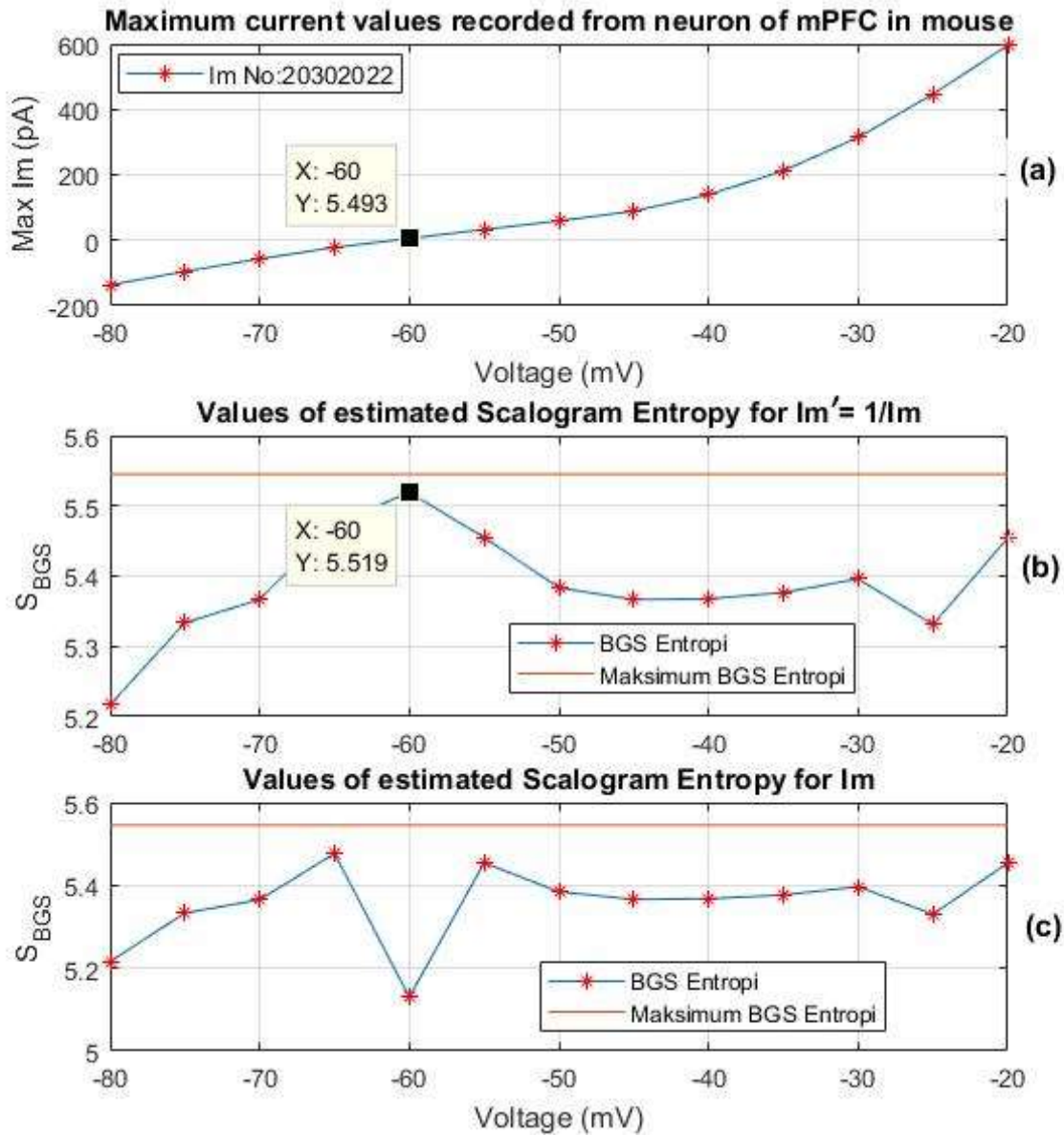

**Figure 5:** Registration file number: 20302022. **(a)** Maximum membrane current values recorded from neuron of the mPFC that were evoked in the range of –80 mV to –20 mV. The voltage value closest to the Nernst equilibrium potential of ions is -60 mV. **(b)** The Scalogram entropy graph for membrane potential changes. Values of the Scalogram entropy of the inverse of the current time series $I'_m = \frac{1}{I_m}$. **(c)** Values of Scalogram entropy of the current time series $I_m$.

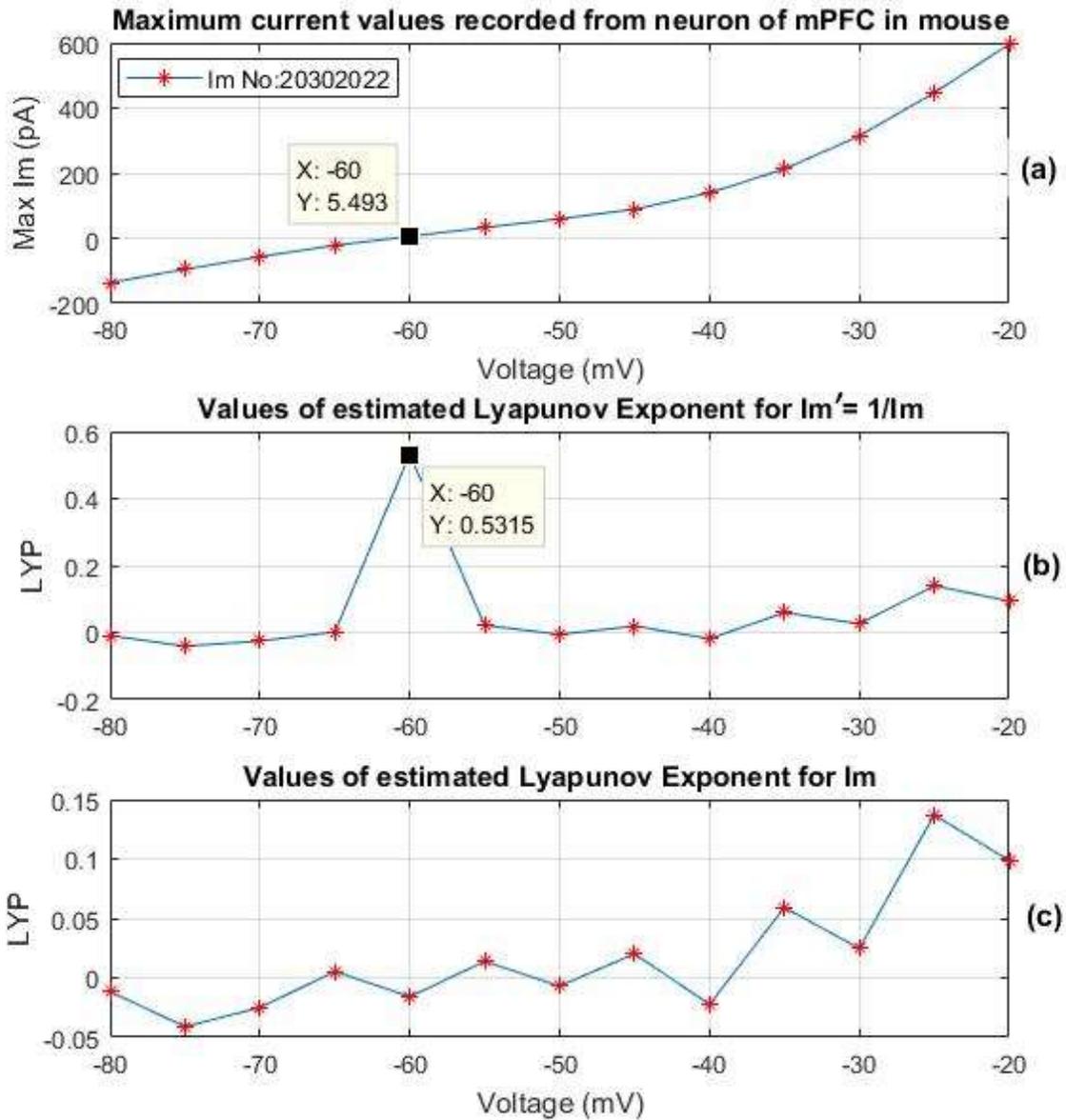

**Figure 6:** Registration file number: 20302022. **(a)** Maximum membrane current values recorded from neuron of the mPFC that were evoked in the range of –80 mV to –20 mV. The ionic current signal approaches a value of zero at a potential value of -60 mV (where the equilibrium potential of ions is noted). **(b)** The largest Lyapunov exponent graph for membrane potential changes. Values of the largest Lyapunov exponent of the inverse of the current time series $I'_m = \frac{1}{I_m}$. **(c)** Values of the largest Lyapunov exponent of the current time series $I_m$.

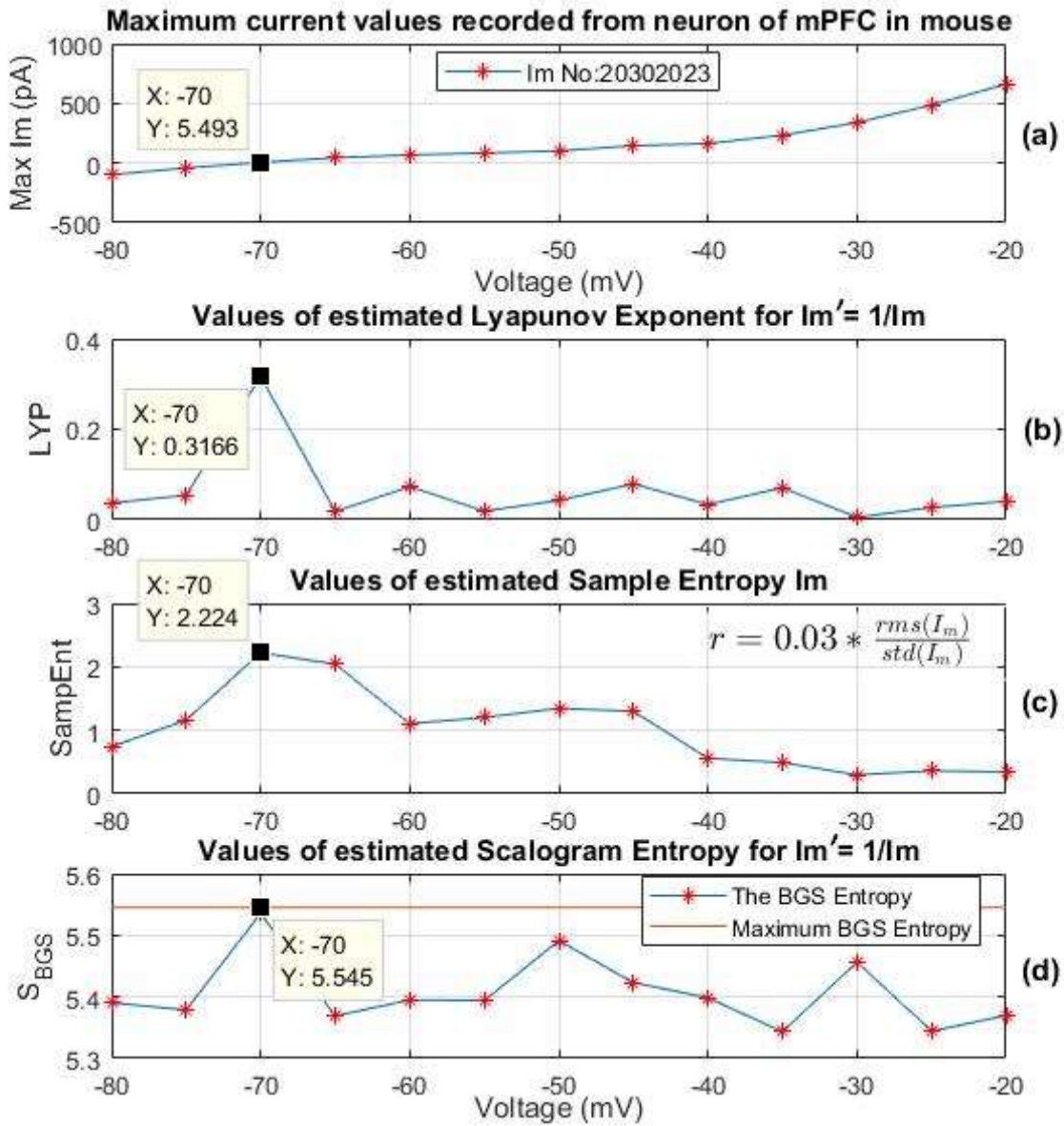

**Figure 7:** Registration file number: 20302023. Graphics of the entropy and largest Lyapunov exponent for membrane potential changes. **(a)** Maximum membrane current values recorded from neuron of the mPFC that were evoked in the range of –80 mV to –20 mV. The ionic current signal approaches a value of zero at a potential value of -70 mV (where the equilibrium potential of ions is noted). **(b)** Values of the largest Lyapunov exponent of the inverse of the current time series $I'_m = \frac{1}{I_m}$. **(c)** Sample entropy values calculated for $r = 0.03 * \frac{rms(I_m)}{std(I_m)}$. **(d)** Values of the Scalogram entropy of the inverse of the current time series $I'_m = \frac{1}{I_m}$.

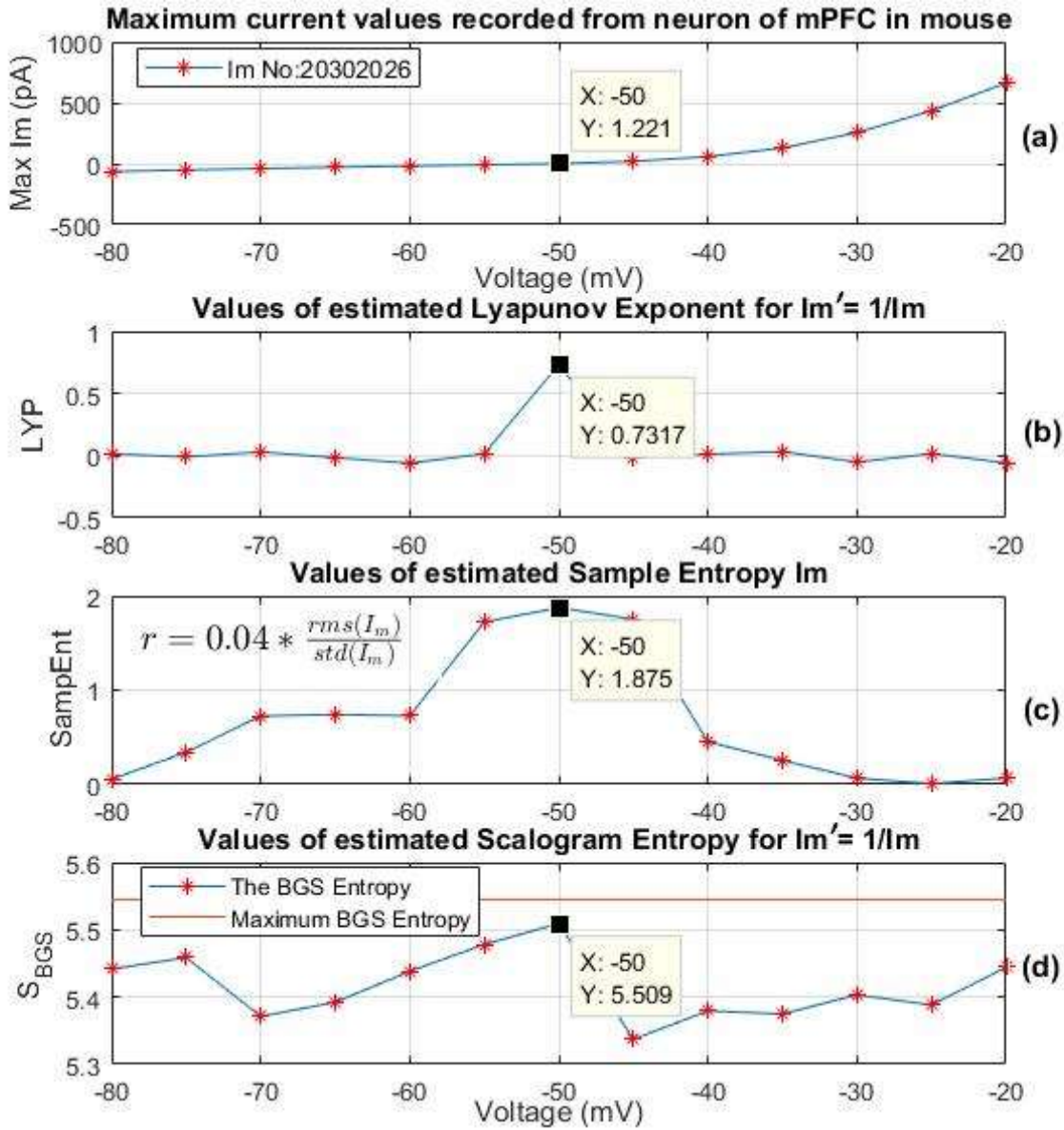

**Figure 8:** Registration file number: 20302026. Graphics of the entropy and largest Lyapunov exponent for membrane potential changes. **(a)** Maximum membrane current values recorded from neuron of the mPFC that were evoked in the range of –80 mV to –20 mV. The ionic current signal approaches a value of zero at a potential value of -50 mV (where the equilibrium potential of ions is noted). **(b)** Values of the largest Lyapunov exponent of the inverse of the current time series $I'_m = \frac{1}{I_m}$. **(c)** Sample entropy values calculated for $r = 0.04 * \frac{rms(I_m)}{std(I_m)}$. **(d)** Values of the Scalogram entropy of the inverse of the current time series $I'_m = \frac{1}{I_m}$.

### 3.2. Testing the assumptions in this study on Logistic map

To test the reliability of the above assumptions, they were also applied to the Logistic map given by the difference equation (16). The entropy and maximum Lyapunov exponent spectrum graphs of the logistic map were plotted by using the proposed assumptions. These spectrum graphs should be similar to the classical solutions of the logistic map.

$$X_{n+1} = AX_n(1 - X_n) \qquad (16)$$

**Figure** 9 shows bifurcation diagrams of the Logistic map for $X_{n+1}$, $\frac{1}{X_{n+1}}$ and $\frac{rms(X_{n+1})}{X_{n+1}}$, plotted in the interval of $3.5 \leq A \leq 4$. Where, $X_{n+1}$ is the time series obtained from the numerical solutions of the Logistic map (16), $rms(X_{n+1})$ is root mean square (rms) of the series of $X_{n+1}$. The chaotic and periodic regions of these bifurcation diagrams do not change as seen in **Figure** 9 (a)-(c).

**Figure** 10 (b) shows Sample entropy spectrum of the Logistic map $X_{n+1}$ for $r = 0.2 * std(X_{n+1})$.
**Figure** 10 (c) shows Sample entropy spectrum of the Logistic map $X_{n+1}$ for $r = 0.02 * \frac{rms(X_{n+1})}{std(X_{n+1})}$.
**Figure** 10 (d) shows Sample entropy spectrum of the Logistic map which is $\frac{1}{X_{n+1}}$ for $r = 0.01 * \frac{rms(1/X_{n+1})}{std(1/X_{n+1})}$. The results of Sample entropy spectrums are similar to each other as seen in **Figure** 10 (b)-(d).

**Figure** 11 shows Scalogram entropy spectrums of the Logistic map for $X_{n+1}$, $\frac{1}{X_{n+1}}$ and $\frac{rms(X_{n+1})}{X_{n+1}}$ respectively. Scalogram entropy spectrums of $\frac{1}{X_{n+1}}$ and $\frac{rms(X_{n+1})}{X_{n+1}}$ give similar results to the spectrum of $X_{n+1}$ as seen in **Figure** 11 (b)-(d).

**Figure** 12 shows Lyapunov exponent spectrums of the Logistic map for $X_{n+1}$, $\frac{1}{X_{n+1}}$ and $\frac{rms(X_{n+1})}{X_{n+1}}$ respectively. Lyapunov exponent spectrums of $\frac{1}{X_{n+1}}$ and $\frac{rms(X_{n+1})}{X_{n+1}}$ give similar results to the spectrum of $X_{n+1}$ as seen in **Figure 12** (b)-(d). Here, the largest Lyapunov exponents were not calculated from the derivative of equation of the logistic map (16). The algorithm proposed (15) for calculating the largest Lyapunov exponents of time series $X_{n+1}$ was preferred.

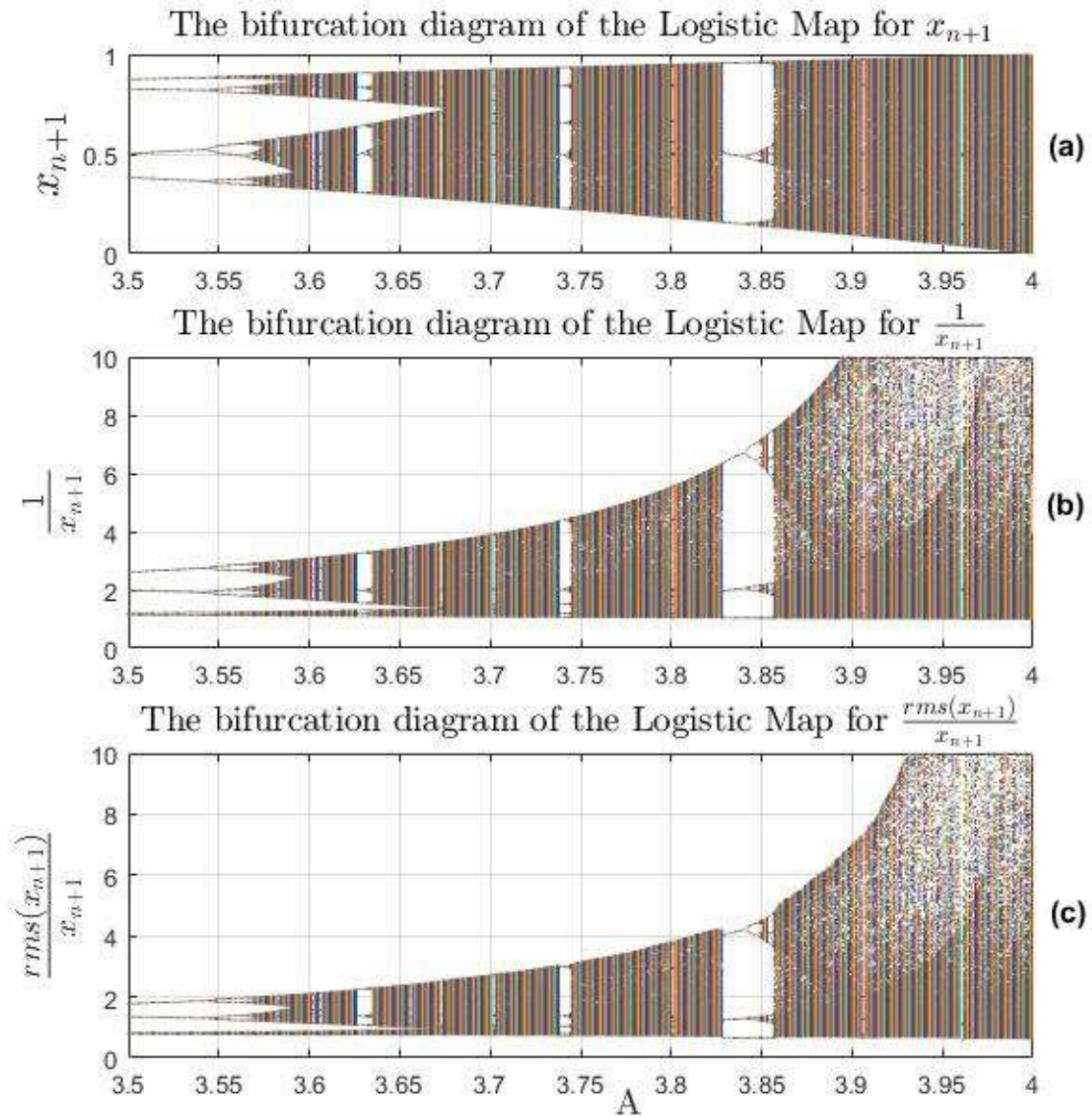

**Figure 9:** Bifurcation diagrams of the Logistic map. **(a)** The bifurcation diagram of $X_{n+1}$. **(b)** The bifurcation diagram of $\frac{1}{X_{n+1}}$. **(c)** The bifurcation diagram of $\frac{rms(X_{n+1})}{X_{n+1}}$.

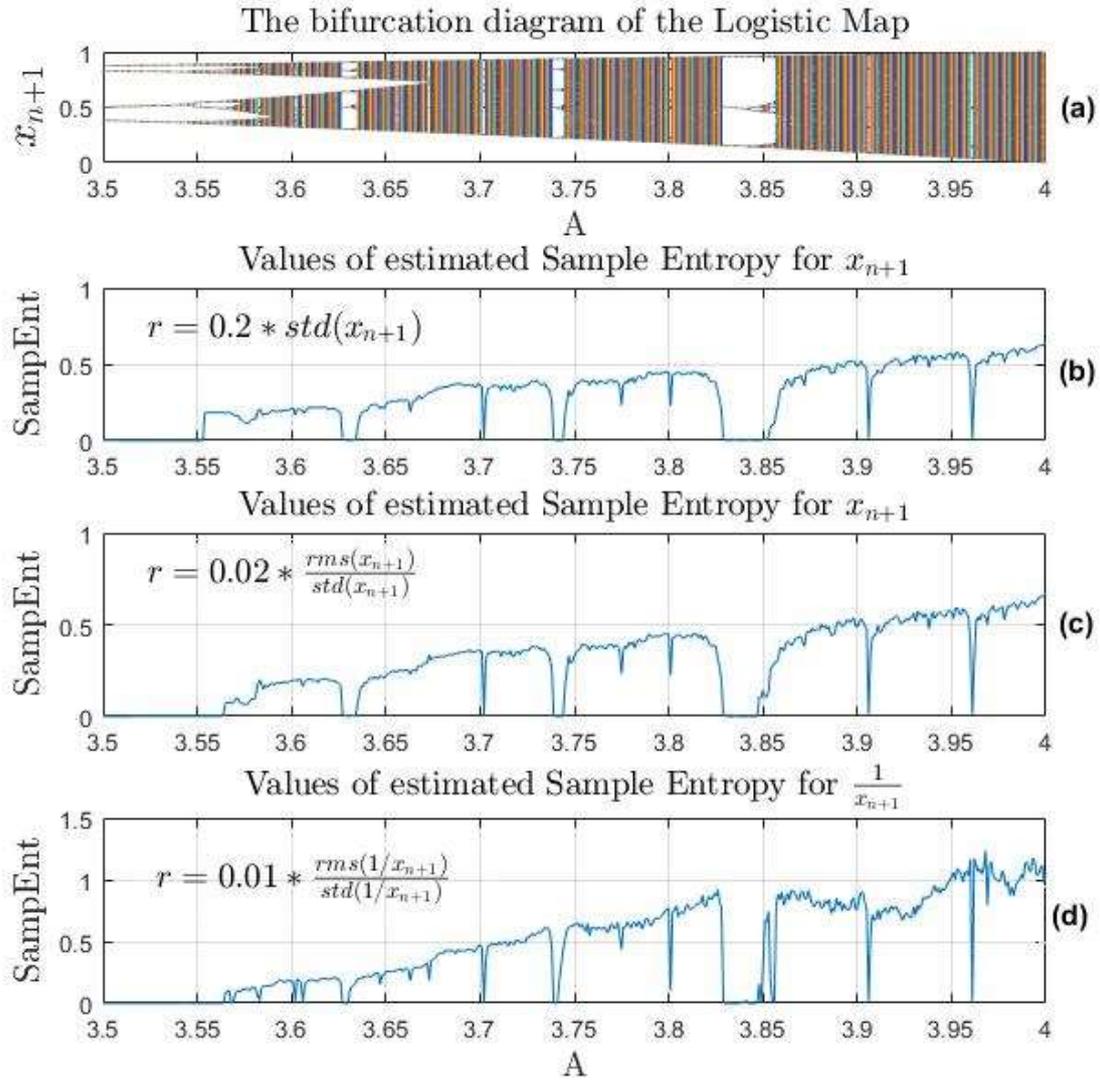

**Figure 10: (a)** The bifurcation diagram of the Logistic map $X_{n+1}$. **(b)** The sample entropy spectrum of $X_{n+1}$ for $r = 0.2 * std(X_{n+1})$. **(c)** The sample entropy spectrum of $X_{n+1}$ for $r = 0.02 * \frac{rms(X_{n+1})}{std(X_{n+1})}$. **(d)** The sample entropy spectrum of $\frac{1}{X_{n+1}}$ for $r = 0.01 * \frac{rms(1/X_{n+1})}{std(1/X_{n+1})}$.

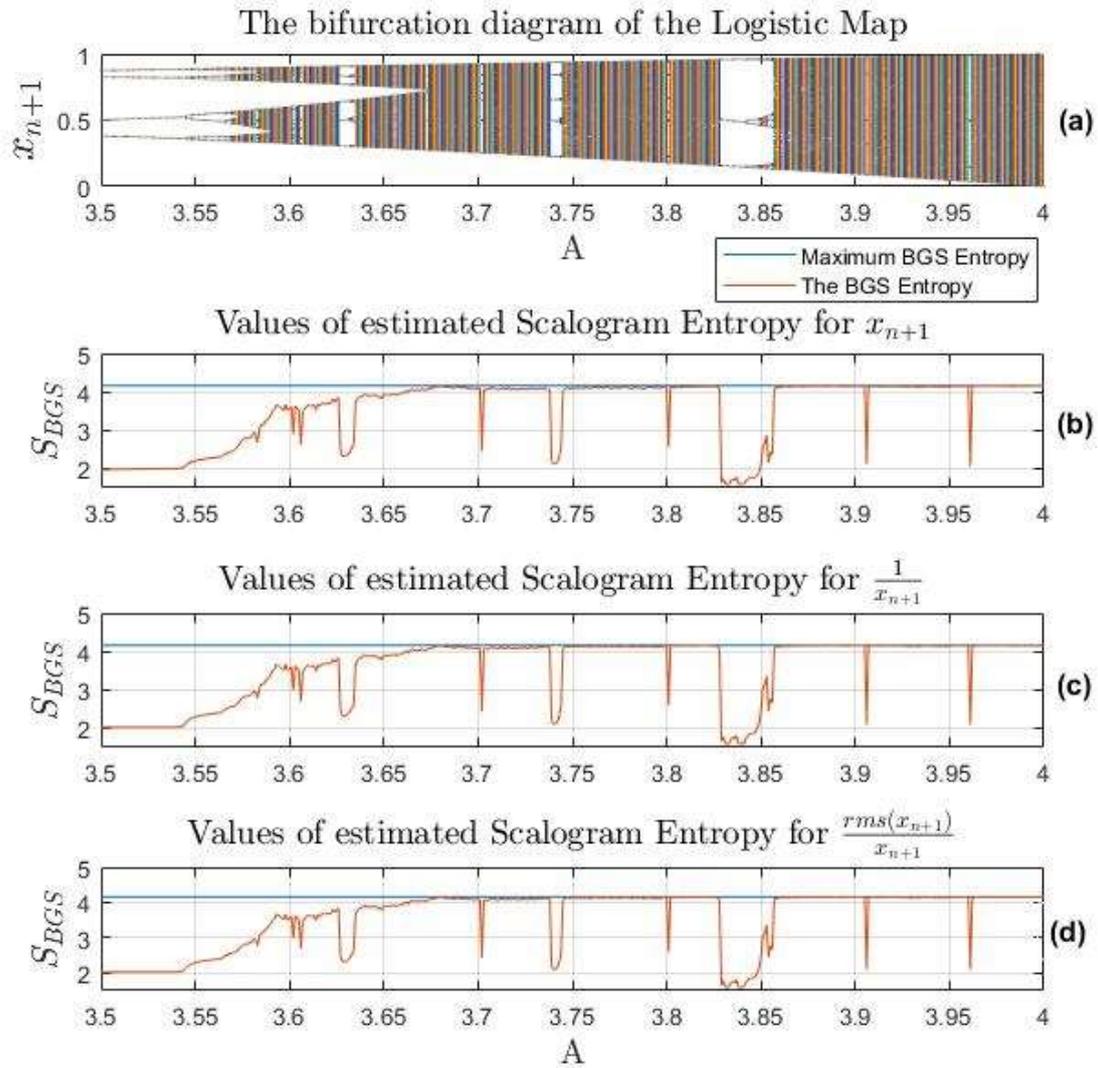

**Figure 11: (a)** The bifurcation diagram of the Logistic map $X_{n+1}$. **(b)** The scalogram entropy spectrum of $X_{n+1}$. **(c)** The scalogram entropy spectrum of $\frac{1}{X_{n+1}}$. **(d)** The scalogram entropy spectrum of $\frac{rms(X_{n+1})}{X_{n+1}}$.

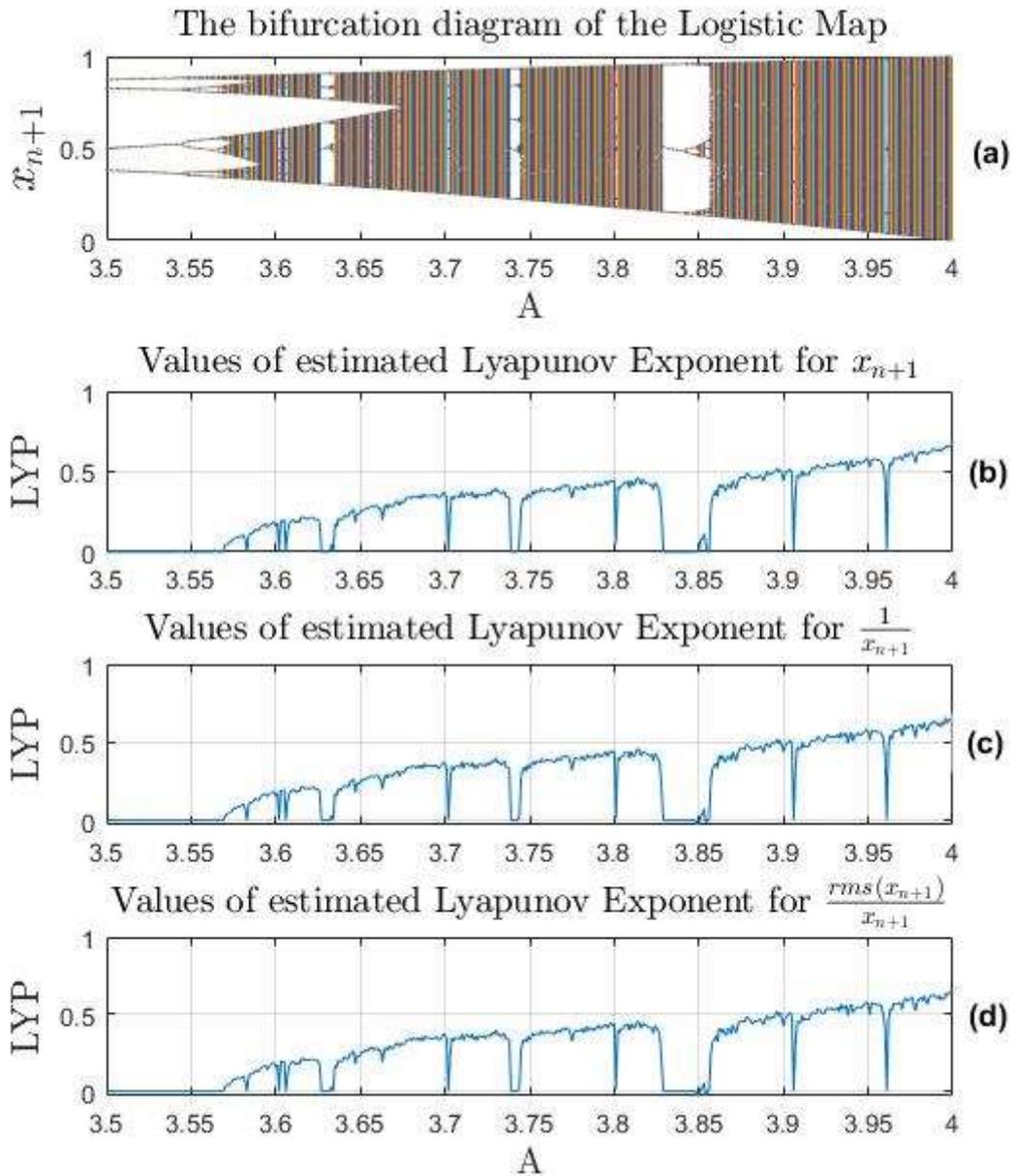

**Figure 12: (a)** The bifurcation diagram of the Logistic map $X_{n+1}$. **(b)** The maximum Lyapunov exponent spectrum of $X_{n+1}$. **(c)** The maximum Lyapunov exponent spectrum of $\frac{1}{X_{n+1}}$. **(d)** The maximum Lyapunov exponent spectrum of $\frac{rms(X_{n+1})}{X_{n+1}}$.

## 4. Discussion and Conclusions

In this study, the entropy of a neuron cell was calculated by using membrane current signals, which were recorded at different membrane potential values. Sample entropy and scalogram entropy were used as measurement methods. It is well known that entropy increases as the system approaches equilibrium. This is why the entropy will be at its maximum at the equilibrium. Similarly, the entropy of a living cell will be maximised at the Nernst equilibrium potential of the ion currents in the process of electrical activity of the cell. This assumption was used to calibrate the accuracy of the calculated entropy. Three mouse mPFC neurons with different equilibrium potentials were selected in an experiment carried out previously [39]. In this experiment, time series of neuronal ionic currents were recorded using the whole-cell patch clamp technique with a voltage clamp protocol in the range -80 mV to -20 mV. The membrane current signals of the mPFC neurons are shown in **Figure** 2.

As seen in **Figures** 4(c) and 5(c), in the traditional calculations made for the entropies of the Sample and Salogram, it was observed that the entropy values did not peak around the Nernst equilibrium potential of the ionic currents. Similarly, **Figure** 6(c) shows that the largest Lyapunov exponent value does not peak around the equilibrium. These results are not consistent with the general assumption that entropy is maximal at equilibrium. In order to provide that these results are consistent with the general assumption, some modifications were made to the measurement methods.

Firstly; for the Sample entropy calculation of cell ionic current signals, It was proposed that the value of the tolerance $r$ can be determined according to the formula $r \equiv c * \frac{rms(data)}{std(data)}$. Here, the coefficient $c$ is adjusted in such a way that the entropy peaks around the equilibrium potential of the ionic current. In calculations made according to the proposed tolerance (13), the Sample entropy values peak around the Nernst equilibrium potential of the ionic currents as seen in the **Figures** 4(b), 7(c) and 8(c).

Secondly; as a result of the trials, in the calculations made for the inverse of the ionic current time series $I'_m = \frac{1}{I_m}$, it was observed that both the Scalogram entropy and the largest Lyapunov exponent values peaked at the Nernst equilibrium potential of the ionic currents, as seen in the **Figures** 5(b), 6(b), 7(b), 7(d), 8(b) and 8(d). Here, the approach of using the inverse of the time series rather than itself produces the expected results.

Additionally, the reliability of both the proposed tolerance (13) for the Sample entropy and the approach of using the inverse of the time series instead of itself for the Scalogram entropy and the Lyapunov exponent were tested on the Logistic map. The **Figures** 10, 11 and 12 show that the obtained solutions give similar results to the classical solutions. Therefore, these approaches can be used.

The results show that the behaviour of living cells, which changes with membrane voltage, can be monitored by entropy measurements (i.e., the graph of the entropy change depending on the membrane potential). This method can also be used to detect differences in the behaviour of tumour and normal cells or the effects of drugs on cells.

**CRediT authorship contribution statement** Single author

**Funding** This research was not funded.

**Data Availability Statement** Data will be made available on reasonable request.

**Conflict of interest** The author has no Conflict of interest to disclose

**References**

[1] C. Beisbart & S. Hartmann (eds), Probabilities in Physics, Oxford University Press 2011, 115–42.
[2] Wehrl, Alfred. "General properties of entropy." Reviews of Modern Physics 50.2 (1978): 221.
[3] Greven, Andreas, Gerhard Keller, and Gerald Warnecke, eds. Entropy. Vol. 47. Princeton University Press, 2014.
[4] Jaynes, Edwin T. "Gibbs vs Boltzmann entropies." American Journal of Physics 33.5 (1965): 391-398.
[5] Penrose, O. Foundations of Statistical Mechanics: A Deductive Treatment; Pergamon: Oxford, UK, (1970).
[6] Schrodinger, E; "What is life?", Cambridge University Press, UK, 1994.
[7] Harkin, Emerson F., et al. "Temporal derivative computation in the dorsal raphe network revealed by an experimentally driven augmented integrate-and-fire modeling framework." Elife 12 (2023): e72951
[8] Richman, Joshua S., and J. Randall Moorman. "Physiological time-series analysis using approximate entropy and sample entropy." American journal of physiology-heart and circulatory physiology 278.6 (2000): H2039-H2049.
[9] Akıllı, Mahmut, Nazmi Yılmaz, and K. Gediz Akdeniz. "Study of the q-Gaussian distribution with the scale index and calculating entropy by normalized inner scalogram." Physics Letters A 383.11 (2019): 1099-1104.
[10] Akıllı, Mahmut, Nazmi Yılmaz, and K. Gediz Akdeniz. "The 'wavelet'entropic index q of non-extensive statistical mechanics and superstatistics." Chaos, Solitons & Fractals 150 (2021): 111094.
[11] Akıllı, Mahmut, and Nazmi Yılmaz. "Entropy of the quantum fluctuations of fermionic instantons in the Universe." Modern Physics Letters A 37.16 (2022): 2250101.
[12] Alberts, Bruce. Molecular biology of the cell. Garland science, 2017.
[13] Wright, Stephen H. "Generation of resting membrane potential." Advances in physiology education 28.4 (2004): 139-142.
[14] Sperelakis, Nicholas, ed. Cell physiology sourcebook: a molecular approach. Elsevier, 2001.
[15] Kulbacka, Julita, et al. "Cell membrane transport mechanisms: Ion channels and electrical properties of cell membranes." Transport across natural and modified biological membranes and its implications in physiology and therapy (2017): 39-58.
[16] Ashrafuzzaman, Mohammad, and Jack A. Tuszynski. Membrane biophysics. Springer Science & Business Media, 2012.
[17] Hille, Bertil. "Ionic channels of excitable membranes, Sinauer Assoc." Inc., Sunderland, MA (2001).
[18] Mārgineanu, D-G. "Entropy changes associated with a nerve impulse." Kybernetik 11.2 (1972): 73-76.
[19] Fang, Xiaona, et al. "Nonequilibrium physics in biology." Reviews of Modern Physics 91.4 (2019): 045004.
[20] Venegas-Gomez, Araceli. "The Thermodynamics of the living organisms: entropy production in the cell." arXiv preprint arXiv:1410.8820 (2014).
[21] Davies, Paul CW, Elisabeth Rieper, and Jack A. Tuszynski. "Self-organization and entropy reduction in a living cell." Biosystems 111.1 (2013): 1-10.
[22] https://commons.wikimedia.org/wiki/File:Basis_of_Membrane_Potential2-en.svg
[23] R. Clausius, ''On the Motive Power of Heat, and on the Laws which may be Deduced from it for the Theory of Heat,'' Ann. Phys. 79, (1850).
[24] Gibbs, Josiah Willard. Elementary principles in statistical mechanics: developed with especial reference to the rational foundations of thermodynamics. C. Scribner's sons, 1902.
[25] Shannon, Claude Elwood. "A mathematical theory of communication." The Bell system technical journal 27.3 (1948): 379-423.

[26] Penrose, Oliver. Foundations of statistical mechanics: a deductive treatment. Courier Corporation, 2005.
[27] Gray, Robert M. Entropy and information theory. Springer Science & Business Media, 2011.
[28] Xiong, Wanting, Luca Faes, and Plamen Ch Ivanov. "Entropy measures, entropy estimators, and their performance in quantifying complex dynamics: Effects of artifacts, nonstationarity, and long-range correlations." Physical review E 95.6 (2017): 062114.
[29] Mallat, Stéphane. A wavelet tour of signal processing. Elsevier, 1999.
[30] Benítez, Rafael, Vicente J. Bolós, and M. E. Ramírez. "A wavelet-based tool for studying non-periodicity." Computers & Mathematics with Applications 60.3 (2010): 634-641.
[31] Bolós, Vicente J., et al. "The windowed scalogram difference: a novel wavelet tool for comparing time series." Applied Mathematics and Computation 312 (2017): 49-65.
[32] Pincus, Steven M. "Approximate entropy as a measure of system complexity." Proceedings of the National Academy of Sciences 88.6 (1991): 2297-2301.
[33] Delgado-Bonal, Alfonso, and Alexander Marshak. "Approximate entropy and sample entropy: A comprehensive tutorial." Entropy 21.6 (2019): 541.
[34] Yılmaz, Nazmi, Mahmut Akıllı, and Mine Ak. "Temporal evolution of entropy and chaos in low amplitude seismic wave prior to an earthquake." Chaos, Solitons & Fractals 173 (2023): 113585.
[35] Sandri, Marco. "Numerical calculation of Lyapunov exponents." Mathematica Journal 6.3 (1996): 78-84.
[36] Kantz, Holger. "A robust method to estimate the maximal Lyapunov exponent of a time series." Physics letters A 185.1 (1994): 77-87.
[37] Wolf, Alan, et al. "Determining Lyapunov exponents from a time series." Physica D: nonlinear phenomena 16.3 (1985): 285-317.
[38] Rosenstein, Michael T., James J. Collins, and Carlo J. De Luca. "A practical method for calculating largest Lyapunov exponents from small data sets." Physica D: Nonlinear Phenomena 65.1-2 (1993): 117-134.
[39] Harkin, Emerson et al. (2023). Patch-clamp recordings from dorsal raphe neurons [Dataset]. Dryad. https://doi.org/10.5061/dryad.66t1g1k2w
[40] Sakmann, Bert, and Erwin Neher. "Patch clamp techniques for studying ionic channels in excitable membranes." Annual review of physiology 46.1 (1984): 455-472.
[41] Neher, Erwin, and Bert Sakmann. "The patch clamp technique." Scientific American 266.3 (1992): 44-51
[42] Sontheimer, Harald. "Whole-cell patch-clamp recordings." Patch-clamp applications and protocols. Totowa, NJ: Humana Press, 1995. 37-73.